\documentstyle[12pt,axodraw]{article}
\title{$W$-boson production at upgraded HERA 
         }
\author{
         M.N.Dubinin \\
       {\small \it Institute of Nuclear Physics, Moscow State University} \\
       {\small \it  119899 Moscow, Russia} \\
         H.S.Song \\
       {\small \it Center for Theoretical Physics,}
       {\small \it Seoul National University} \\
       {\small \it Seoul, 151-742, Korea} }
\date{}

\begin{document}
\maketitle

\begin{abstract}
Event characteristics of $W$ boson production at 
HERA collider are untrivial and sensitive to the production mechanisms.
We analyse the distributions of the four particle final state
defined by the complete set of $W$ producing perturbative
leading order diagrams in the Standard Model and its extension
with the anomalous effective lagrangian in the gauge sector.
\end{abstract}

It is important to understand in details the mechanisms of
$W$-boson production at upgraded HERA. The number of $W$-production 
events will be large enough both in leptonic and hadronic $W$ decay
channels, giving potentially large backgrounds to the signals of new
physics \cite{anomalies}, such as contact interactions, leptoquarks
and $R$-parity conserving or violating SUSY processes. At the
same time the single $W$ production is influenced by $\gamma WW$ and
$ZWW$ vertices that still are not precisely measured. While the weak
current
couplings of gauge bosons are measured with the accuracy of 10$^{-4}$
\cite{weak}, the accuracy of anomalous couplings restriction in the
gauge boson sector from the Tevatron data is around 0.5 (in
units of dimensionless parameters $\kappa$, $\lambda$) only 
(\cite{accuracy}, see also \cite{Schild}), so the self-interaction of
vector bosons is not experimentally fixed at comparable level of precision.

$W$ boson production in the electron-proton mode can be observed in the
main channel $e^- p \to e^- \mu^+ \nu_{\mu} X$ and maybe a few events $e^-
p \to \nu_e \mu^+ \bar \nu_{\mu} X$ can be reconstructed. Positron-proton
mode
$e^+ p \to e^+ \mu^+ \nu_{\mu} X$ has practically the same total rate and 
event characteristics, because the Feynman graph topology of the
positron-proton mode is different only for some $W$ and $Z$ exchange weak
diagrams giving the contributions that are small in comparison with
the dominant $t$-channel photon exchange diagrams.
The complete set of 10 tree level diagrams for the main process of
$W$-boson production $e^- p \to e^- \mu^+ \nu_{\mu} X$ is shown
in Fig.1. In the simpler $2 \to 3$ process approximation \cite{BZ} of 
$W$-boson on-shell $M_{\mu \nu}=(p_{\nu}+p_{\mu})^2=m^2_W$  
we have to keep seven $s$-channel diagrams with an outgoing $W$
boson line and omit ladder diagrams 1,6 and 7. However this  
approximation of infinitely small $W$ width $m_W \Gamma_{tot}/[(M^2_{\mu
\nu}-m^2_W)^2 + m^2_W\Gamma_{tot}^2] = \delta (M_{\mu \nu}-m_W)$, 
being rather satisfactory for the calculation of total 
$W$ production rate, is not sufficient for the analysis of 
some specific features of event topology, like particle 
distributions from the multiperipheral mechanisms.
Taking $W$ boson off-shell $W \to \mu \nu_{\mu}$ in the $2 \to 4$
process approximation requires ladder graphs 1,6,7 in order to
preserve gauge invariance \cite{BVZ}.   

Main contribution to the cross section comes from the diagram 3
containing photon and quark $t$-channel propagators. When the
t-channel gamma and quark are close to  mass shell the QCD 
corrections are large, potentially developing a nonperturbative
regime of 'resolved photon' \cite{resolved}. The kinematical
configurations of resolved photon are usually separated by a cut near the
$u$-channel quark pole $\Lambda^2=-(p_q-p_W)^2$. Resolved $W$-production
is not more than a quark-antiquark fusion where one of the quarks
appears as a constituent of gamma and another one as a constituent of
proton. Existing parametrizations of quark distributions inside
gamma (measured in a relatively low $Q^2$ $\gamma \gamma$ collisions and
then extrapolated to the region of $Q^2\sim m^2_W$), seem
to be not as precise as PDF in proton, so the resolved $W$-production
cannot be calculated with the same reliability \cite{DS} as the
perturbative leading order (also called 'direct') contribution of ten
diagrams in Fig.1. Double counting of the perturbative part of the photon
structure function and the collinear configuration of the direct process 
amplitude is removed by the subtraction of LO $\gamma \to q
q^\prime$ splitting term. Resulting contribution of the
resolved part is of order 10\% of the $W$-production cross section. 

Alternative method of the resolved part separation by a cut near the
$t$-channel gamma
pole $Q^2=-(p_e-p^\prime_e)^2$ \cite{nason} uses a delicate
application of the equivalent gamma approximation to the subprocess
with $t$-channel quark $\gamma q \to q^\prime W^*$. In this
case by definition of \cite{nason} the main 'direct' (or
'photoproduction') contribution comes from a configuration with the
quasireal gamma and electron at a small angle $\theta_e$ with the beam.
The $2 \to 3$ amplitude with $\theta_e \ge$ 5 deg. (seven diagrams of
Fig.1, $W$ on-shell) corresponds to 'DIS contribution'.
Ladder diagrams are omitted. Resolved photon contribution becomes
dominant in the framework of this method. Important role of the
NLO QCD corrections to the resolved part $q^\prime q \to W$, giving
the enhancement of it by about 40\%, is demonstrated by an explicit
calculation. 

In the following we shall discuss
the distributions corresponding to direct (perturbative leading
order) part of the $W$ production amplitude in the Standard Model and beyond
\cite{DS}. Resolved gamma is separated by a $\Lambda^2$ cut.

\unitlength=0.8pt
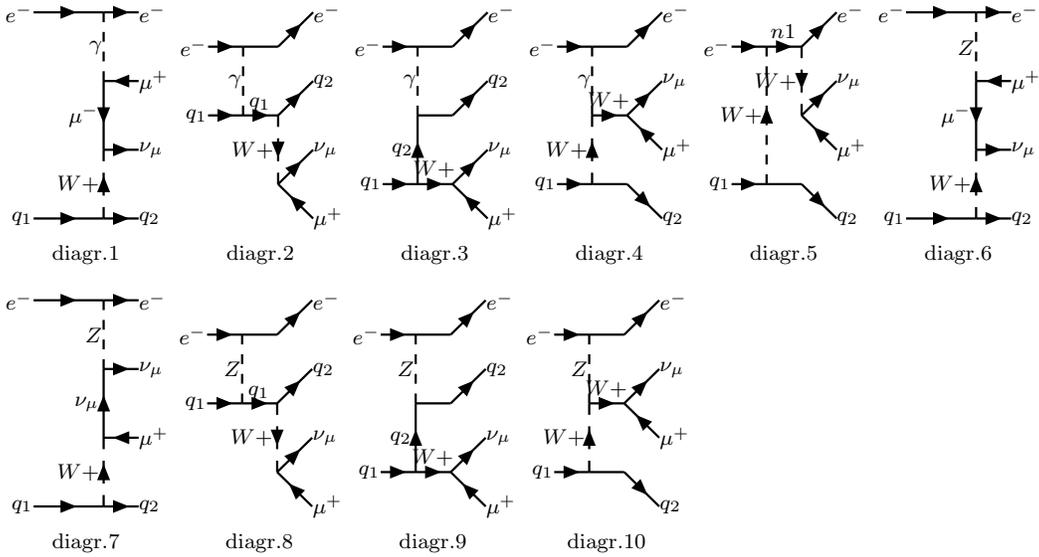
\begin{figure}[h]
{\def\chepscale{1.2} 
\unitlength=\chepscale pt
\SetWidth{0.7}      
\SetScale{\chepscale}
\scriptsize    
\begin{picture}(50,90)(0,0)
\Text(8.2,77.8)[r]{$e^-$}
\ArrowLine(8.5,77.8)(30.5,77.8) 
\Text(41.8,77.8)[l]{$e^-$}
\ArrowLine(30.5,77.8)(41.5,77.8) 
\Text(30.1,67.0)[r]{$\gamma$}
\DashLine(30.5,77.8)(30.5,56.2){3.0} 
\Text(41.8,56.2)[l]{$\mu^+$}
\ArrowLine(41.5,56.2)(30.5,56.2) 
\Text(29.1,45.4)[r]{$\mu^-$}
\ArrowLine(30.5,56.2)(30.5,34.6) 
\Text(41.8,34.6)[l]{$\nu_{\mu}$}
\ArrowLine(30.5,34.6)(41.5,34.6) 
\Text(29.1,23.8)[r]{$W+$}
\DashArrowLine(30.5,13.0)(30.5,34.6){3.0} 
\Text(8.2,13.0)[r]{$q_1$}
\ArrowLine(8.5,13.0)(30.5,13.0) 
\Text(41.8,13.0)[l]{$q_2$}
\ArrowLine(30.5,13.0)(41.5,13.0) 
\Text(25,0)[b] {diagr.1}
\end{picture} \ 
\begin{picture}(50,90)(0,0)
\Text(8.2,67.0)[r]{$e^-$}
\ArrowLine(8.5,67.0)(19.5,67.0) 
\Line(19.5,67.0)(30.5,67.0) 
\Text(41.8,77.8)[l]{$e^-$}
\ArrowLine(30.5,67.0)(41.5,77.8) 
\Text(19.1,56.2)[r]{$\gamma$}
\DashLine(19.5,67.0)(19.5,45.4){3.0} 
\Text(8.2,45.4)[r]{$q_1$}
\ArrowLine(8.5,45.4)(19.5,45.4) 
\Text(24.8,48.2)[b]{$q_1$}
\ArrowLine(19.5,45.4)(30.5,45.4) 
\Text(41.8,56.2)[l]{$q_2$}
\ArrowLine(30.5,45.4)(41.5,56.2) 
\Text(29.1,34.6)[r]{$W+$}
\DashArrowLine(30.5,45.4)(30.5,23.8){3.0} 
\Text(41.8,34.6)[l]{$\nu_{\mu}$}
\ArrowLine(30.5,23.8)(41.5,34.6) 
\Text(41.8,13.0)[l]{$\mu^+$}
\ArrowLine(41.5,13.0)(30.5,23.8) 
\Text(25,0)[b] {diagr.2}
\end{picture} \ 
\begin{picture}(50,90)(0,0)
\Text(8.2,67.0)[r]{$e^-$}
\ArrowLine(8.5,67.0)(19.5,67.0) 
\Line(19.5,67.0)(30.5,67.0) 
\Text(41.8,77.8)[l]{$e^-$}
\ArrowLine(30.5,67.0)(41.5,77.8) 
\Text(19.1,56.2)[r]{$\gamma$}
\DashLine(19.5,67.0)(19.5,45.4){3.0} 
\Line(19.5,45.4)(30.5,45.4) 
\Text(41.8,56.2)[l]{$q_2$}
\ArrowLine(30.5,45.4)(41.5,56.2) 
\Text(18.1,34.6)[r]{$q_2$}
\ArrowLine(19.5,23.8)(19.5,45.4) 
\Text(8.2,23.8)[r]{$q_1$}
\ArrowLine(8.5,23.8)(19.5,23.8) 
\Text(24.8,26.6)[b]{$W+$}
\DashArrowLine(19.5,23.8)(30.5,23.8){3.0} 
\Text(41.8,34.6)[l]{$\nu_{\mu}$}
\ArrowLine(30.5,23.8)(41.5,34.6) 
\Text(41.8,13.0)[l]{$\mu^+$}
\ArrowLine(41.5,13.0)(30.5,23.8) 
\Text(25,0)[b] {diagr.3}
\end{picture} \ 
\begin{picture}(50,90)(0,0)
\Text(8.2,67.0)[r]{$e^-$}
\ArrowLine(8.5,67.0)(19.5,67.0) 
\Line(19.5,67.0)(30.5,67.0) 
\Text(41.8,77.8)[l]{$e^-$}
\ArrowLine(30.5,67.0)(41.5,77.8) 
\Text(19.1,56.2)[r]{$\gamma$}
\DashLine(19.5,67.0)(19.5,45.4){3.0} 
\Text(24.8,48.2)[b]{$W+$}
\DashArrowLine(19.5,45.4)(30.5,45.4){3.0} 
\Text(41.8,56.2)[l]{$\nu_{\mu}$}
\ArrowLine(30.5,45.4)(41.5,56.2) 
\Text(41.8,34.6)[l]{$\mu^+$}
\ArrowLine(41.5,34.6)(30.5,45.4) 
\Text(18.1,34.6)[r]{$W+$}
\DashArrowLine(19.5,23.8)(19.5,45.4){3.0} 
\Text(8.2,23.8)[r]{$q_1$}
\ArrowLine(8.5,23.8)(19.5,23.8) 
\Line(19.5,23.8)(30.5,23.8) 
\Text(41.8,13.0)[l]{$q_2$}
\ArrowLine(30.5,23.8)(41.5,13.0) 
\Text(25,0)[b] {diagr.4}
\end{picture} \ 
\begin{picture}(50,90)(0,0)
\Text(8.2,67.0)[r]{$e^-$}
\ArrowLine(8.5,67.0)(19.5,67.0) 
\Text(24.8,69.8)[b]{$n1$}
\ArrowLine(19.5,67.0)(30.5,67.0) 
\Text(41.8,77.8)[l]{$e^-$}
\ArrowLine(30.5,67.0)(41.5,77.8) 
\Text(29.1,56.2)[r]{$W+$}
\DashArrowLine(30.5,67.0)(30.5,45.4){3.0} 
\Text(41.8,56.2)[l]{$\nu_{\mu}$}
\ArrowLine(30.5,45.4)(41.5,56.2) 
\Text(41.8,34.6)[l]{$\mu^+$}
\ArrowLine(41.5,34.6)(30.5,45.4) 
\Text(18.1,45.4)[r]{$W+$}
\DashArrowLine(19.5,23.8)(19.5,67.0){3.0} 
\Text(8.2,23.8)[r]{$q_1$}
\ArrowLine(8.5,23.8)(19.5,23.8) 
\Line(19.5,23.8)(30.5,23.8) 
\Text(41.8,13.0)[l]{$q_2$}
\ArrowLine(30.5,23.8)(41.5,13.0) 
\Text(25,0)[b] {diagr.5}
\end{picture} \ 
\begin{picture}(50,90)(0,0)
\Text(8.2,77.8)[r]{$e^-$}
\ArrowLine(8.5,77.8)(30.5,77.8) 
\Text(41.8,77.8)[l]{$e^-$}
\ArrowLine(30.5,77.8)(41.5,77.8) 
\Text(30.1,67.0)[r]{$Z$}
\DashLine(30.5,77.8)(30.5,56.2){3.0} 
\Text(41.8,56.2)[l]{$\mu^+$}
\ArrowLine(41.5,56.2)(30.5,56.2) 
\Text(29.1,45.4)[r]{$\mu^-$}
\ArrowLine(30.5,56.2)(30.5,34.6) 
\Text(41.8,34.6)[l]{$\nu_{\mu}$}
\ArrowLine(30.5,34.6)(41.5,34.6) 
\Text(29.1,23.8)[r]{$W+$}
\DashArrowLine(30.5,13.0)(30.5,34.6){3.0} 
\Text(8.2,13.0)[r]{$q_1$}
\ArrowLine(8.5,13.0)(30.5,13.0) 
\Text(41.8,13.0)[l]{$q_2$}
\ArrowLine(30.5,13.0)(41.5,13.0) 
\Text(25,0)[b] {diagr.6}
\end{picture} \ 
\begin{picture}(50,90)(0,0)
\Text(8.2,77.8)[r]{$e^-$}
\ArrowLine(8.5,77.8)(30.5,77.8) 
\Text(41.8,77.8)[l]{$e^-$}
\ArrowLine(30.5,77.8)(41.5,77.8) 
\Text(30.1,67.0)[r]{$Z$}
\DashLine(30.5,77.8)(30.5,56.2){3.0} 
\Text(41.8,56.2)[l]{$\nu_{\mu}$}
\ArrowLine(30.5,56.2)(41.5,56.2) 
\Text(29.1,45.4)[r]{$\nu_{\mu}$}
\ArrowLine(30.5,34.6)(30.5,56.2) 
\Text(41.8,34.6)[l]{$\mu^+$}
\ArrowLine(41.5,34.6)(30.5,34.6) 
\Text(29.1,23.8)[r]{$W+$}
\DashArrowLine(30.5,13.0)(30.5,34.6){3.0} 
\Text(8.2,13.0)[r]{$q_1$}
\ArrowLine(8.5,13.0)(30.5,13.0) 
\Text(41.8,13.0)[l]{$q_2$}
\ArrowLine(30.5,13.0)(41.5,13.0) 
\Text(25,0)[b] {diagr.7}
\end{picture} \ 
\begin{picture}(50,90)(0,0)
\Text(8.2,67.0)[r]{$e^-$}
\ArrowLine(8.5,67.0)(19.5,67.0) 
\Line(19.5,67.0)(30.5,67.0) 
\Text(41.8,77.8)[l]{$e^-$}
\ArrowLine(30.5,67.0)(41.5,77.8) 
\Text(19.1,56.2)[r]{$Z$}
\DashLine(19.5,67.0)(19.5,45.4){3.0} 
\Text(8.2,45.4)[r]{$q_1$}
\ArrowLine(8.5,45.4)(19.5,45.4) 
\Text(24.8,48.2)[b]{$q_1$}
\ArrowLine(19.5,45.4)(30.5,45.4) 
\Text(41.8,56.2)[l]{$q_2$}
\ArrowLine(30.5,45.4)(41.5,56.2) 
\Text(29.1,34.6)[r]{$W+$}
\DashArrowLine(30.5,45.4)(30.5,23.8){3.0} 
\Text(41.8,34.6)[l]{$\nu_{\mu}$}
\ArrowLine(30.5,23.8)(41.5,34.6) 
\Text(41.8,13.0)[l]{$\mu^+$}
\ArrowLine(41.5,13.0)(30.5,23.8) 
\Text(25,0)[b] {diagr.8}
\end{picture} \ 
\begin{picture}(50,90)(0,0)
\Text(8.2,67.0)[r]{$e^-$}
\ArrowLine(8.5,67.0)(19.5,67.0) 
\Line(19.5,67.0)(30.5,67.0) 
\Text(41.8,77.8)[l]{$e^-$}
\ArrowLine(30.5,67.0)(41.5,77.8) 
\Text(19.1,56.2)[r]{$Z$}
\DashLine(19.5,67.0)(19.5,45.4){3.0} 
\Line(19.5,45.4)(30.5,45.4) 
\Text(41.8,56.2)[l]{$q_2$}
\ArrowLine(30.5,45.4)(41.5,56.2) 
\Text(18.1,34.6)[r]{$q_2$}
\ArrowLine(19.5,23.8)(19.5,45.4) 
\Text(8.2,23.8)[r]{$q_1$}
\ArrowLine(8.5,23.8)(19.5,23.8) 
\Text(24.8,26.6)[b]{$W+$}
\DashArrowLine(19.5,23.8)(30.5,23.8){3.0} 
\Text(41.8,34.6)[l]{$\nu_{\mu}$}
\ArrowLine(30.5,23.8)(41.5,34.6) 
\Text(41.8,13.0)[l]{$\mu^+$}
\ArrowLine(41.5,13.0)(30.5,23.8) 
\Text(25,0)[b] {diagr.9}
\end{picture} \ 
\begin{picture}(50,90)(0,0)
\Text(8.2,67.0)[r]{$e^-$}
\ArrowLine(8.5,67.0)(19.5,67.0) 
\Line(19.5,67.0)(30.5,67.0) 
\Text(41.8,77.8)[l]{$e^-$}
\ArrowLine(30.5,67.0)(41.5,77.8) 
\Text(19.1,56.2)[r]{$Z$}
\DashLine(19.5,67.0)(19.5,45.4){3.0} 
\Text(24.8,48.2)[b]{$W+$}
\DashArrowLine(19.5,45.4)(30.5,45.4){3.0} 
\Text(41.8,56.2)[l]{$\nu_{\mu}$}
\ArrowLine(30.5,45.4)(41.5,56.2) 
\Text(41.8,34.6)[l]{$\mu^+$}
\ArrowLine(41.5,34.6)(30.5,45.4) 
\Text(18.1,34.6)[r]{$W+$}
\DashArrowLine(19.5,23.8)(19.5,45.4){3.0} 
\Text(8.2,23.8)[r]{$q_1$}
\ArrowLine(8.5,23.8)(19.5,23.8) 
\Line(19.5,23.8)(30.5,23.8) 
\Text(41.8,13.0)[l]{$q_2$}
\ArrowLine(30.5,23.8)(41.5,13.0) 
\Text(25,0)[b] {diagr.10}
\end{picture} \ 
}
\caption{Feynman diagrams for the process $e^- q_1 \rightarrow
         e^- \mu^+ \nu_{\mu} q_2$}
\end{figure}
  
Calculation of the leading order amplitude (ten diagrams in Fig.1)
was performed by means of CompHEP package \cite{CompHEP}. We used
the 'overall' prescription \cite{KPS,BVZ} for vector boson propagators.
Electron mass was kept nonzero for the regularization of $t$-channel
gamma pole. Numerical level cancellation of the $1/Q^4$ gamma propagator
pole in the separately taken squared diagrams to the $1/Q^2$ pole in the
entire amplitude, required by the gauge invariance, can be demonstrated
explicitly \cite{DS}.
 
We show the event characteristics of the process
$e^- p \to e^- \mu^+ \nu_{\mu} X$ in Figs.2-4 (proton structure
functions MRS A \cite{MRS}).
First row of plots in each figure represents the distributions of the
electron, second row of plots - distributions of the (anti)muon, and third
row of plots shows the distributions of the final quark (all
partonic subprocesses contributing to $W$ production are summed).
Different cuts for $t$-channel quark momentum $\Lambda^2=-(p_q- p_W)^2$
were imposed for the calculation of event characteristics in Fig.2
and Fig.3, where $\Lambda$ is equal to 0.2 GeV and 5 GeV, correspondingly.
The distributions in Fig.4 were calculated at $\Lambda=$ 5 GeV with
the phase space cut 20 GeV around the $W$ pole: 
$M_W-20$ GeV $\le M(\mu \nu_{\mu}) \le M_W+20$ GeV. This cut is
used in the EPVEC generator \cite{BVZ}, so the distributions in
Fig.4 are calculated using the phase space cuts similar to the cuts
used in \cite{noyes}. Additional phase space cuts in the EPVEC that
were introduced to ensure the numerical stability of the muon
pole integration in the ladder diagram 1, Fig.1,
lead to the EPVEC total rate regularly smaller than the CompHEP total
rate, where the muon pole region is exactly integrated (see Table 1).

\begin{table}[t]
\begin{center}
\begin{tabular}{|c|c|c|c|c|}
\hline
 & \multicolumn{2}{|c|}{EPVEC } &
\multicolumn{2}{|c|}{CompHEP} \\ \hline
 & $e^+ p$ & $e^- p$ & $e^+ p$ & $e^- p$ \\ \hline
        \multicolumn{5}{|c|}{$\Lambda$ cut 0.2 GeV, no $W$ pole cut} \\
\hline
MRS A  & 71.7 & 70.8 & 93.1 & 92.1    \\
CTEQ4L & 71.5  & 70.8 & 93.0 & 91.9   \\ \hline
        \multicolumn{5}{|c|}{$\Lambda$ cut 5 GeV, no $W$ pole cut} \\
\hline
MRS A  & 50.7 & 49.5 & 68.9 & 68.6   \\
CTEQ4L & 50.9 & 49.9 & 69.1 & 68.9   \\ \hline
        \multicolumn{5}{|c|}{$\Lambda$ cut 5 GeV, $W$ pole cut $\pm$20
GeV} \\ \hline
MRS A  & 44.2 & 43.2 & 53.6 & 51.7   \\
CTEQ4L & 44.5 & 43.4 & 53.7 & 51.9   \\ \hline
\end{tabular}
\end{center}
\caption{EPVEC and CompHEP total rates (fb) for the processes $e^+ p \to
e^+ \mu^+ \nu_{\mu} X$ and $e^- p \to e^- \mu^+ \nu_{\mu} X$ Direct part
only, proton structure functions MRS A \cite{MRS} and CTEQ4L \cite{CTEQ}.
Parton distribution functions factorisation scale $m_W$, resolved gamma
region is separated by a $\Lambda$ cut.
}
\end{table}
   
Soft muons in the distributions $d\sigma/dE_{\mu}$ and $d\sigma/dp_{T \mu}$
come from the ladder diagrams 1,6,7 in Fig.1. Jets at
the angle 180 degrees with the proton beam appear from diagram 3,
Fig.1, when the quasireal photon produces a quark-antiquark pair collinear to
the initial electron. 
We can see that the backscattered jet peak is sensitive to
the value of $\Lambda$ cut and gradually goes down as
we increase it. But this decrease of direct
contribution should be compensated by the 'photon remnant' in the 
contribution from the resolved part. Muons at 180 degree with the proton beam
from the ladder diagrams appear at any value of $\Lambda$ - only the shape
of $d\sigma / dE_{\mu}$ is slightly affected and the normalization
is changed. Kinematical cut around the $W$-pole implemented
in 'canonical' EPVEC removes soft muons and muons at 180 degree with the
beam completely. It is interesting to notice that the four-fermion final
state configurations with backscattered soft muons are observed
in the simulation for LEP2 \cite{LEP2}, where in the channel $e^+ e^-
\to e^- \bar \nu_e \mu^+ \nu_{\mu}$ diagram topologies are the
same (if we replace the quark line by the electron line).

In the muonic channels the total
cross section is equal approximately to 150-160 fb, giving about
35 events/year at the integrated luminosity 200 $pb^{-1}$.
Additional kinematical cuts are necessary in the electron channels for
separation of misidentification backgrounds, and the number of
identifiable events from $W \rightarrow e \nu_e$ is slightly smaller.
In total at upgraded HERA collider it could be possible to observe 
about 60 $W$-production events/year. At present time HERA luminosity is
too small to produce a number of $W$ sufficient for a quantitative analysis.
Presently available H1 data
\cite{H1} shows one event in the electron channel (2.4 events expected)
and two events in the muon channel (0.8 events expected), with the topology
compatible with $W$ production processes (three more H1 events in the muon
channel have the kinematic properties different from those given by $W$
production mechanisms). ZEUS data \cite{ZEUS} shows three events in the
electron channel and none of the events passes through kinematical cuts
in the muon channel. Detailed discussion can be found in \cite{Spiesb}.

Anomalous U(1) gauge invariant, C and P parity conserving effective
lagrangian in the gauge sector can be taken in the form \cite{HPZH}

\begin{equation}
L_{eff}=g_V (W^+_{\mu \nu} W^{\mu} V^{\nu}
            -W^{\mu \nu} W^+_{\mu} V_{\nu}
            +\kappa \; W^+_{\mu} W_{\nu} V^{\mu \nu}
  +\frac{\lambda}{m^2_W} W^+_{\rho \mu} W^{\mu}_{\; \; \nu} V^{\nu \rho})
\end{equation}
where $g_{\gamma}=e$ and $g_Z=e cos\vartheta_W/sin\vartheta_W$,
$W_{\mu \nu}=\partial_{\mu} W_{\nu}- \partial_{\nu} W_{\mu}$,
$V_{\mu \nu}=\partial_{\mu} V_{\nu}- \partial_{\nu} V_{\mu}$.
and
$\lambda$, $\kappa$ are dimensionless parameters. Following the tradition
the value of $\lambda$ is expressed in the units of $m_W$.
We show one of the distributions
calculated using the anomalous effective lagrangian (1) in Fig.5.

Taking into account some typical detector experimental cuts nesessary
for the separation of misidentification backgrounds and
realistic experimental acceptancies \cite{noyes} we can estimate the
following limits \cite{DS} for
$\Delta \kappa$ and $\lambda$, giving the observable deviation of
total cross section from the Standard Model value at 68\% and 95\%
confidence level:
\begin{eqnarray*}
-1.70 < \lambda < 1.70,\phantom{xxxx} -1.05 < \Delta \kappa < 0.48,
                       \phantom{xxxx}  68\% \quad CL \\ 
-2.24 < \lambda < 2.24,\phantom{xxxxxxxxxxxx} \Delta \kappa < 0.89,
                       \phantom{xxxx}  95\% \quad CL
\end{eqnarray*}
at the integrated luminosity 200 $pb^{-1}$, and
\begin{eqnarray*}
-1.03 < \lambda < 1.03,\phantom{xxxx} -0.31  <\Delta \kappa < 0.27,
                       \phantom{xxxx}  68\% \quad CL \\
-1.75 < \lambda < 1.75,\phantom{xxxx} -0.58  <\Delta \kappa < 0.46,
                       \phantom{xxxx}  95\% \quad CL
\end{eqnarray*}
at the integrated luminosity 1000 $pb^{-1}$. The limits from Tevatron
collider available at present time \cite{accuracy} are close to the possible
upgraded HERA limits at the luminosity 1000 $pb^{-1}$.

M.D. is grateful to M.Spira, M.Kuze and D.Zeppenfeld for useful
discussions and he thanks very much C.Diaconu and D.Waters for help in
the comparison of EPVEC and CompHEP results.

\newpage

\unitlength 1.0cm

\begin{figure}
\begin{picture}(17,17)
\put(-1,0){\epsfxsize=17cm
         \epsfysize=17 cm \leavevmode \epsfbox{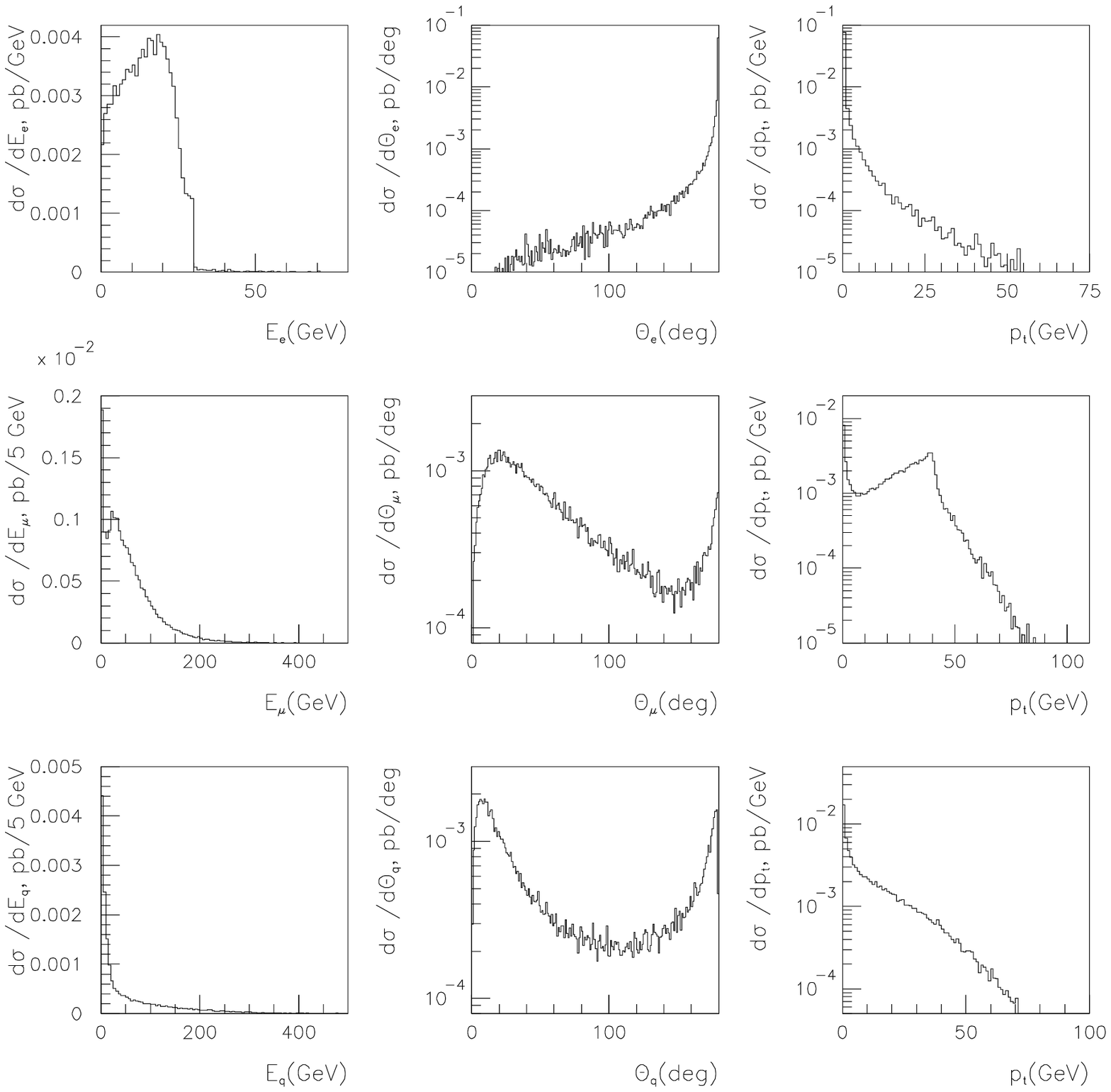}}
\end{picture}
\caption{First row of plots - distributions of the electron energy,
scattering angle and transverse momentum in the process
$e^- p \rightarrow e^- \mu^+ \nu_{\mu} X$. Second row of plots -
distributions of the muon energy, muon scattering angle and transverse
momentum. Third row of plots - distributions of the quark energy, angle
and transverse momentum for the same process. Total cross section
92.1 fb, $t$-channel quark cut $\Lambda=$ 0.2 GeV.}
\end{figure}

\newpage

\begin{figure}
\begin{picture}(17,17)
\put(-1,0){\epsfxsize=17cm
         \epsfysize=17 cm \leavevmode \epsfbox{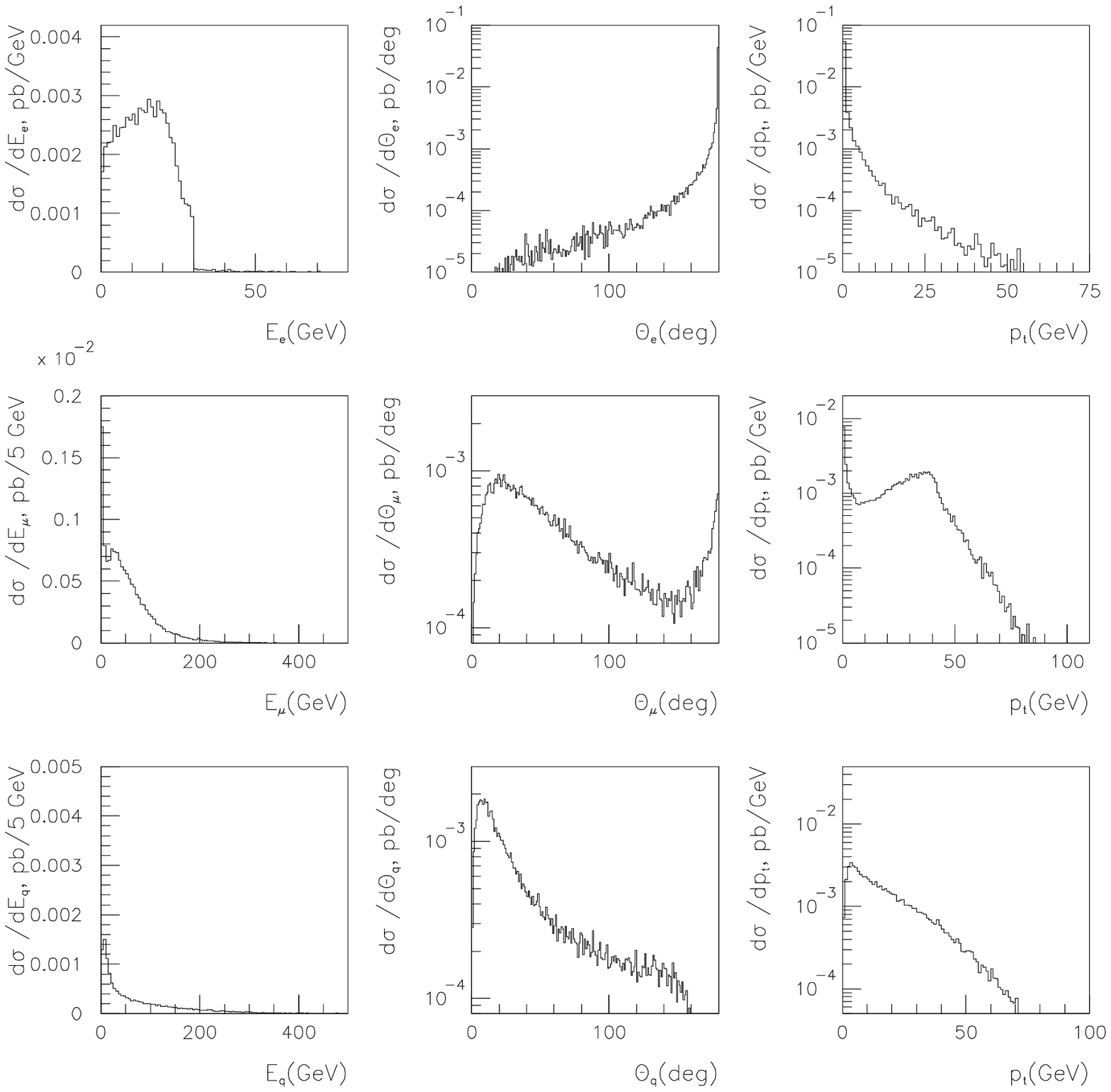}}
\end{picture}
\caption{First row of plots - distributions of the electron energy,
scattering angle and transverse momentum in the process
$e^- p \rightarrow e^- \mu^+ \nu_{\mu} X$. Second row of plots -
distributions of the muon energy, muon scattering angle and transverse
momentum. Third row of plots - distributions of the quark energy, angle
and transverse momentum for the same process. Total cross section
68.6 fb, $t$-channel quark cut $\Lambda=$ 5 GeV.}
\end{figure}

\newpage

\begin{figure}
\begin{picture}(17,17)
\put(-1,0){\epsfxsize=17cm
         \epsfysize=17 cm \leavevmode \epsfbox{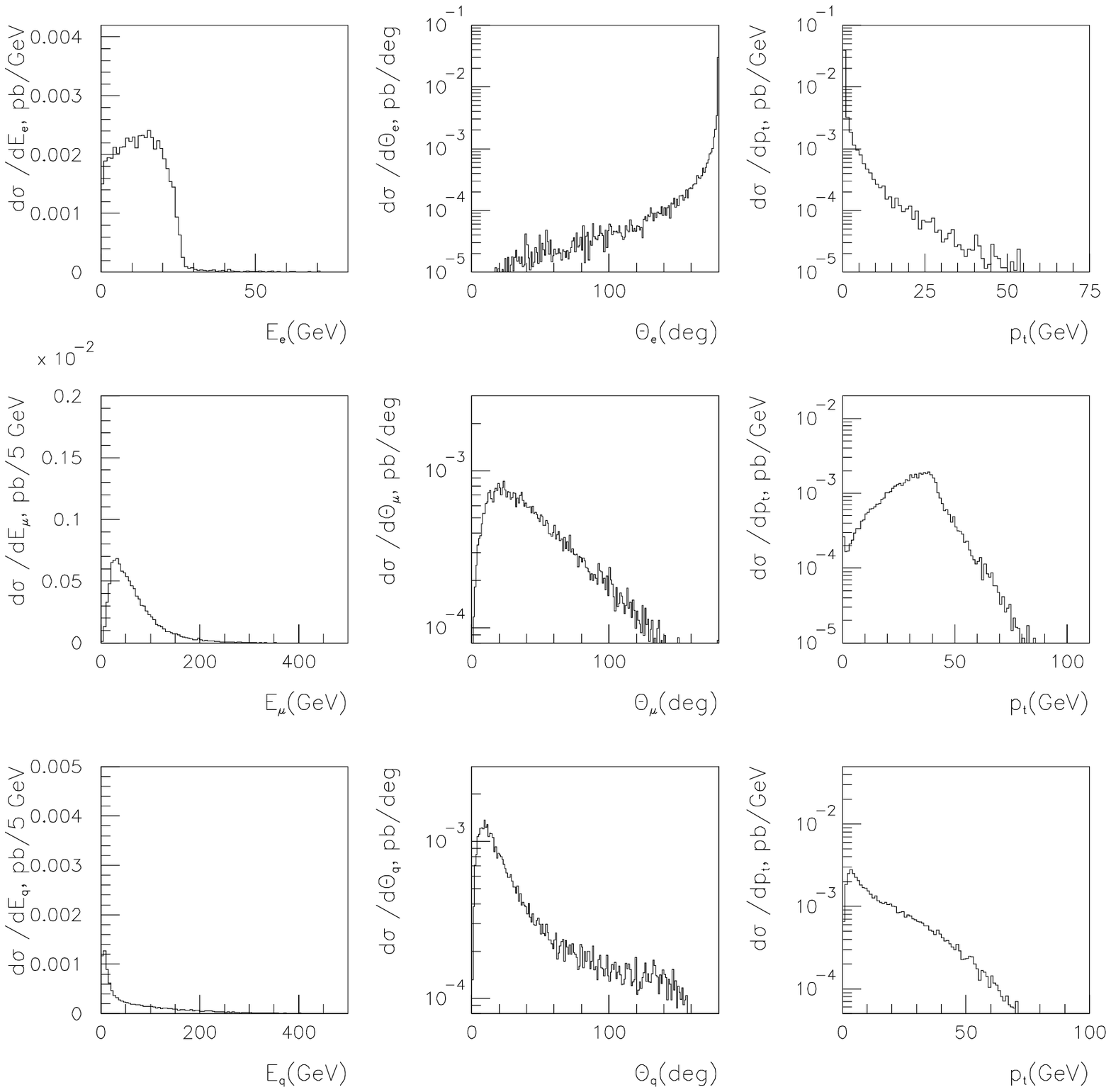}}
\end{picture}
\caption{First row of plots - distributions of the electron energy,
scattering angle and transverse momentum in the process
$e^- p \rightarrow e^- \mu^+ \nu_{\mu} X$. Second row of plots -
distributions of the muon energy, muon scattering angle and transverse
momentum. Third row of plots - distributions of the quark energy, angle
and transverse momentum for the same process. Total cross section
51.7 fb, $t$-channel quark cut $\Lambda=$ 5 GeV, the cut around the $W$
pole 20 GeV, to be compared with
the same distributions obtained by means of EPVEC generator, see [14]}
\end{figure}

\begin{figure}
\begin{picture}(15,15)
\put(-6,-3){\epsfxsize=19cm
         \epsfysize=25 cm \leavevmode \epsfbox{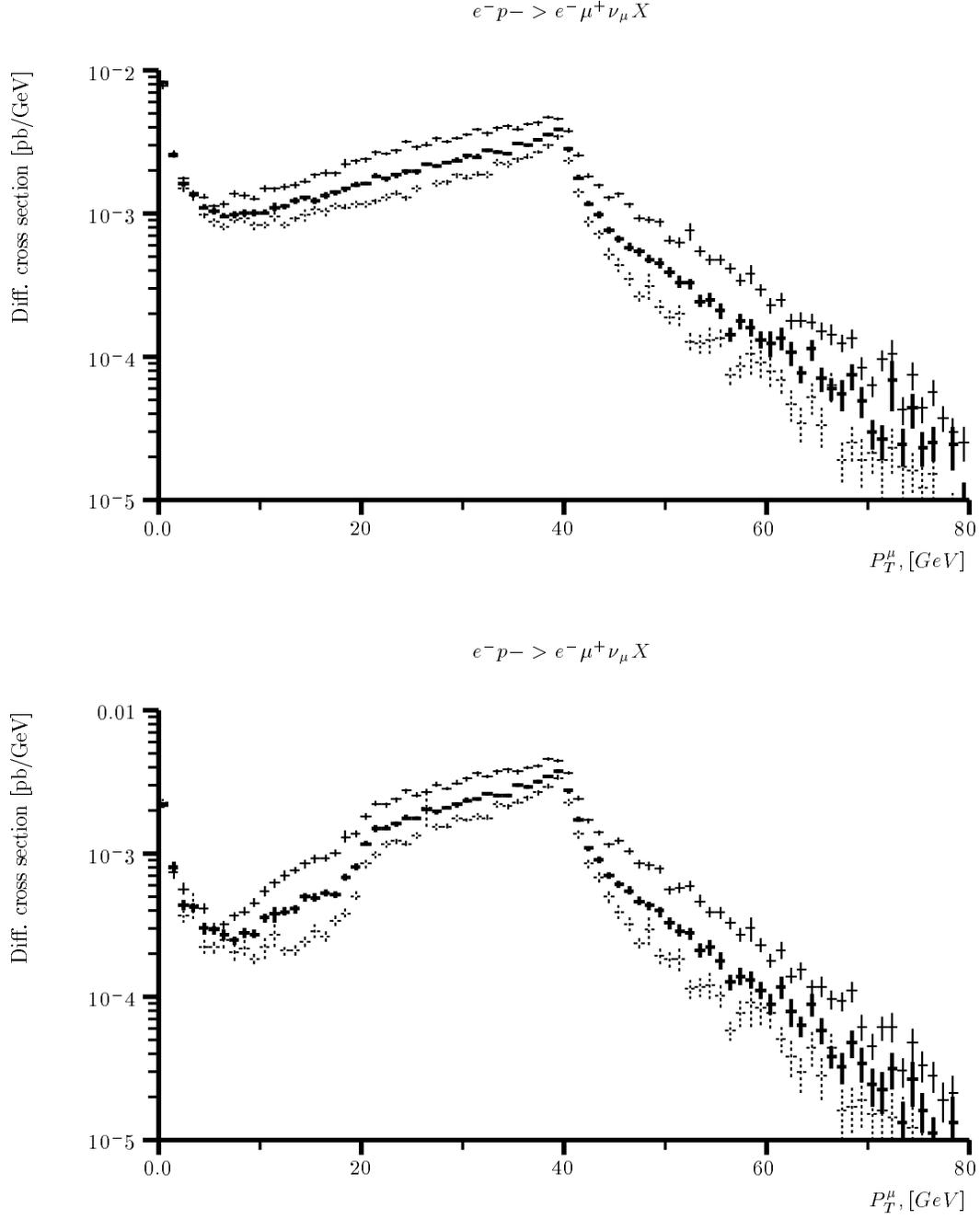}}
\end{picture}
\caption{Distribution of muon transverse momentum in the reaction 
$e^- p \rightarrow e^- \mu^+ \nu_{\mu} X$.
Upper plot: no kinematical cuts, solid lines - standard case, dash
lines - anomalous three vector boson couplings case, $\lambda=0$,
$\kappa=0$, thin solid lines - $\lambda=0$, $\kappa=2$.
Lower plot: the same distributions after the detector kinematical cuts
$E_{\mu} \ge$ 10 GeV, missing $p_T \ge$ 20 GeV.}
\end{figure}

\end{document}